# A review and evaluation of secondary school accountability in England: Statistical strengths, weaknesses, and challenges for 'Progress 8'

Lucy Prior

Centre for Multilevel Modelling

School of Education

University of Bristol, UK

John Jerrim

Department of Quantitative Social Science

UCL Institute of Education, UK

Dave Thomson

FFT Education Datalab

London, UK

George Leckie

Centre for Multilevel Modelling

School of Education

University of Bristol, UK

**Funding**

This research was funded by UK Economic and Social Research Council (ESRC) grant ES/R010285/1 (Prior and Leckie) and ES/T003677/1 (Jerrim).

**Acknowledgements**

This work contains statistical data from Office for National Statistics (ONS), UK, which is Crown Copyright. The use of the ONS statistical data in this work does not imply the endorsement of the ONS in relation to the interpretation or analysis of the statistical data.

**A review and evaluation of secondary school accountability in England: Statistical strengths, weaknesses, and challenges for 'Progress 8'**

## Abstract

School performance measures are published annually in England to hold schools to account and to support parental school choice. This article reviews and evaluates the 'Progress 8' secondary school accountability system for state-funded schools. We assess the statistical strengths and weaknesses of Progress 8 relating to: choice of pupil outcome attainment measure; potential adjustments for pupil input attainment and background characteristics; decisions around which schools and pupils are excluded from the measure; presentation of Progress 8 to users, choice of statistical model, and calculation of statistical uncertainty; and issues related to the volatility of school performance over time, including scope for reporting multi-year averages. We then discuss challenges for Progress 8 raised by the COVID-19 pandemic. Six simple recommendations follow to improve Progress 8 and school accountability in England.

## 1. Introduction

School performance measures are calculated annually in England for all state-funded schools. These measures play a pivotal role in the school accountability system. They are also published online in tables (https://www.compare-school-performance.service.gov.uk/) to hold schools publicly accountable and to assist parents choosing schools (Burgess et al., 2019; West & Pennell, 2000; West, 2010). Republication by the media leads to wider public scrutiny and debate (e.g., The Times, 2020a). School performance measures also play a central role in national discussions about the state of education in England including the performance of different regions, school types and pupil groups (Leckie & Goldstein, 2019; BBC News, 2020a).

From 1992 to 2016, the headline measure of school performance was the percentage of pupils achieving five or more GCSEs (age 16, academic year 11) at A* to C grade (5A*-C) (Leckie & Goldstein, 2017). From 2006, this was redefined to include A*-C passes in both GCSE English and maths. The 5A*-C metric aimed to measure the average *attainment* in each school at the end of Key Stage 4 (KS4) and compulsory secondary schooling. The fundamental criticism of 5A*-C was that it ignored that pupils start secondary schooling with very different Key Stage 2 (KS2) test results (age 11, academic year 7). Therefore, schools' 5A*-C results said more about school differences in pupil prior attainment at intake than they did about school differences in the effectiveness or quality of teaching (Mortimore et al., 1988; Sammons et al., 1997; Teddlie & Reynolds, 2000; Wilson, 2004).

In contrast, Progress 8, the current headline measure of school performance introduced in 2016, aims to measure the average *progress* or improvement in attainment seen in each school over the course of secondary schooling (academic years 7 to 11) (Department for Education, 2020a). In doing so, Progress 8 aims to account for school differences in pupil prior attainment at intake, and hence is widely viewed as a fairer measure for comparing schools for accountability and choice purposes (CooperGibson Research, 2017). While the Government has published various school value-added measures since 2005, which have been used by Ofsted to inform school inspections, these did not feature prominently in the performance tables or public discourse (Leckie & Goldstein, 2017). Now, with Progress 8, the Government has made a school progress measure its headline measure for the first time. Progress 8 therefore represents a radical overhaul of the school accountability system and a fundamental shift in Government thinking, long called for by academic research (Goldstein, 1997; Raudenbush & Willms, 1995; Teddlie & Reynolds, 2000).

Progress 8, however, represents just one possible school progress measure and its underlying methodology and implementation reflect a combination of data and statistical modelling decisions. These decisions have important real-world consequences as they directly impact on schools' scores and therefore the high-stakes judgements made about them. Despite this, there is little academic research on Progress 8. In this article, we review Progress 8, evaluate its statistical strengths and weaknesses, and make simple recommendations to improve it as a measure of school performance and accountability.

The COVID-19 pandemic hugely disrupted school education, leading schools to close nationally in both March 2020 and January 2021 and pupils to lose out on months of learning with many pupils spending additional time out of school self-isolating (Department for Education, 2020b). The 2020 and 2021 GCSE examinations and KS2 tests were cancelled (Department for Education, 2020c; Department for Education, 2020d) and the Government has therefore chosen not to calculate Progress 8 for these two years (Department for Education, 2020e). We therefore also discuss statistical challenges for school accountability and Progress 8 triggered by COVID-19.

While our focus is on statistical concerns with Progress 8, it is essential to acknowledge broader long-standing concerns with the way school performance data has been used to inform school accountability in England over the last three decades. Common criticisms are that the strong link between school performance measures and school accountability generates perverse incentives and unintended consequences detrimental to student learning including: off-rolling disadvantaged pupils, narrowing of the curriculum, teaching to the test, and undue teacher and pupil stress (Amrein-Beardsley, 2014; Foley & Goldstein, 2012; NAHT, 2018; OECD, 2008). Reflecting in part these concerns, Wales, Scotland, and Northern Ireland do not publish school performances tables and have less data driven school accountability systems, although school average exam results are still widely available via the media (Belfast Telegraph, 2019; The Scottish Sun, 2020; WalesOnline, 2019).

The article proceeds as follows. In Section 2, we review the calculation, presentation, and interpretation of Progress 8 in the school performance tables and we illustrate this for a single local authority. In Section 3, we assess the suitability of Attainment 8 as the output measure at KS4 from which Progress 8 scores are derived. In Section 4, we review KS2 scores as the sole input measure to Progress 8 and debate the potential inclusion of pupil background characteristics. In Section 5, we discuss issues surrounding schools and pupils which are excluded from Progress 8. In Section 6, we evaluate the presentation of Progress 8, the choice of statistical model, and the communication of statistical uncertainty. In Section 7, we consider issues connected to the volatility of Progress 8 scores over time. In Section 8, we turn our attention to statistical challenges for Progress 8 triggered by COVID-19. In Section 9 we present six simple recommendations to improve Progress 8 and school accountability in England.

## 2. Calculation, presentation and interpretation of Progress 8

Progress 8 is described as a pupil and school measure of the academic *progress* pupils make over the five years of compulsory secondary schooling (Department for Education, 2020a). A pupil's Progress 8 score is calculated as the difference between their Attainment 8 score at the end of KS4 and the average Attainment 8 scores among all pupils nationally who had the same prior attainment as measured by pupils' average KS2 test scores. A school's Progress 8 score is the average of their pupils' Progress 8 scores and so measures whether pupils in each school, in general, outscore similar pupils nationally on Attainment 8 (Department for Education, 2020a). Thus, positive, zero, and negative scores are interpreted as schools where pupils are learning more rapidly, similarly, or less rapidly than pupils nationally. See the Supplementary Material for further explanation of the Progress 8 methodology.

As KS2 and Attainment 8 scores play key roles in the calculation of Progress 8, it is helpful to discuss both in more detail. KS2 tests are typically taken by all state school pupils at the end of primary school, with students assessed on English (reading, grammar, punctuation, and spelling) and mathematics (arithmetic and reasoning). For Progress 8, pupils are assigned into one of 34 prior attainment groups based on their average fine grade across the KS2 reading and maths tests. Attainment 8 is the main pupil and school measure of *attainment* at the end of compulsory secondary schooling and was introduced in 2016 alongside Progress 8. A pupil's Attainment 8 score is their total score across eight subject qualifications, where each qualification is assigned a score corresponding to the 9-1 grade system (GCSEs still using the old A*-G grade system are assigned a score mapping onto the 9-1 grade system) (Department for Education, 2020a). The eight subject qualification slots are: English and maths (double weighted to reflect the priority the Government places on these subjects); three further subjects that count in the English Baccalaureate (Ebacc: a set of subjects deemed by the Government to stand pupils in good stead for future study and career options (Department for Education, 2019); and three additional subjects not already counted (the ‘Open’ slots). The Ebacc subjects are English, maths, the sciences, the humanities (geography or history), and a language (Department for Education, 2020a). Thus, for example, a pupil with an Attainment 8 score of 90 achieved eight grade 9 GCSEs. A school’s Attainment 8 score is the average of their pupils’ Attainment 8 scores.

As an illustrative example, Table 1 reproduces the Government 2018/19 school performance table published by the Department for Education (DfE) for all mainstream secondary schools in Bristol. The schools are sorted by their Progress 8 scores from highest to lowest and it is this public ranking which leads these tables to be colloquially referred to as ‘school league tables’. The highest scoring school in Bristol was Redland Green School with a Progress 8 score of 0.53: pupils in this school scored higher on Attainment 8 than other pupils nationally who started with the same KS2 scores by, on average, 0.53 GCSE grades per subject.

Progress 8 scores aim to measure the academic performance or effectiveness of each school. In the school value-added modelling literature (Goldstein, 1997; Raudenbush & Willms, 1995), these ‘observed’ scores are viewed as reflecting a combination of each school’s unobserved ‘true’ performance of interest and sampling variation attributed to the unobserved idiosyncratic characteristics of the pupils who happened to attend, and which would not replicate had the school taught different pupils. For school accountability purposes, interest typically lies in using the observed scores to make inferences (statistical statements) about schools’ unobserved true performances. Progress 8 scores are therefore presented with 95% confidence intervals to convey a range of plausible values for the true performance of each school and to test whether the true performance differs from 0 (the performance of the average school). Put differently, the 95% confidence intervals give a sense of the degree to which a school’s observed Progress 8 score would be expected to vary across different random samples of pupils. The smaller the number of pupils, the wider the confidence interval and so the less confident we can be as to the true performance of the school and the more the observed Progress 8 score would be expected to vary from sample to sample.

Table 1 shows the Steiner Academy Bristol has a very wide 95% confidence interval of -1.12 to 0.47 (a range of 1.59 GCSE grades per subject) which includes a score of 0, the score associated with performing in line with the national average. Only 10 pupils are included in the measure for this school; the data do not provide enough information to declare the true performance of the school statistically different from average, despite its low score of -0.32. In contrast, the 95% confidence interval for Redland Green School where 197 pupils are included ranges from 0.35 to 0.72 (a narrow range of 0.37 GCSE grades per subject) and so does not include a score of 0. The 95% confidence interval is sufficiently narrow to be statistically confident that the true performance of the school is above average and therefore that other cohorts of pupils who might equally have attended this school would also have outperformed similar pupils nationally.

The DfE also present one of five Progress 8 bandings for each school calculated as a joint function of its Progress 8 score and 95% confidence interval. The bandings are given the following qualitative descriptions: ‘Well above average’ (about 14% of schools in England in 2019), ‘Above average’ (17%), ‘Average’ (37%), ‘Below average’ (20%), and ‘Well below average’ (12%). The bandings are colour coded to visually convey the status of the school: dark green, light green, yellow, orange, red. To be described as ‘Well above average’, a school must have a Progress 8 score of 0.5 or greater and have the lower end of their 95% confidence interval sitting above zero. Hence, Redland Green School is reported in Table 1 as ‘Well above average’. Definitions of the other bandings are given in Table 1.

## 3. Choice of pupil outcome attainment measure

### Recognising all students, not just borderline students

A consistent critique of 5A*-C was that it incentivised schools to focus excessively on children at the GCSE grade C/D borderline at the expense of more able pupils (Burgess, et al., 2005; West, 2010). Attainment 8, in contrast, is a continuous measure and so all grades, no matter where on the scale, contribute to the overall score. Therefore, a strength of Progress 8 is that it incentivises schools to focus on all children. It is not clear, however, whether the Attainment 8 scale holds equal meaning at all points. Is, for example, the effort required to move pupils between a 4 and a 5 the same as between an 8 and a 9? To the extent to which there are differences, Progress 8 may still generate incentives to concentrate on pupils at specific points in the distribution. Burgess & Thomson (2020) find some limited evidence suggesting that the introduction of Progress 8 shifted the incentive to focus on borderline pupils to lower-attaining pupils.

**Increased control and commonality over subjects studied**

The introduction of Attainment 8 and Progress 8 followed a series of other changes in the accountability system recommended by the Wolf Review of vocational education (Wolf, 2011). In particular, the review recommended a reduction in the range of qualifications eligible to be included in school performance measures, arguing that students in some schools were being entered for qualifications that were not sufficiently ‘rigorous’, offering limited potential for future education and work (Department for Education, 2015). In response, Attainment 8 is heavily weighted (70:30) in favour of Ebacc subjects. Specifically, five of the eight subjects must be Ebacc subjects and two of these, English and Maths, are double weighted.

Schools appear to have responded to the emphasis placed on EBacc subjects now reinforced by Attainment 8 and Progress 8. Entry patterns between 2010/11 and 2015/16 show rising entries for science and humanity GCSEs and an increasing proportion of students taking at least three EBacc subjects, though languages have not shown similar increases (Gill, 2017). This emphasis on EBacc subjects is important to reflect on given subject choice can have significant consequences for student's future lives, both academic and in the labour market (Iannelli, 2013; Moulton, et al., 2018).

This increased commonality in the subjects entered by students across schools should make gaming Progress 8 harder than it was for 5A*-C. The incentive to enter pupils for vocational 'easy' non-GCSE options has been removed. A notable example is the European Computer Driving Licence (ECDL). Investigations revealed that in some schools it was being taught in as little as three-days as a 'fast-track' qualification, leading to its removal from the list of eligible qualifications (Schools Week, 2018). Analysis comparing the performance of schools before and after the change to the ECDL status showed that schools which had entered most of their students for the ECDL tended to see their Progress 8 scores decline following its removal (FFT Education Datalab, 2018a).

Despite the prescription of subjects in Progress 8 increasing commonality to a degree, the mix of subjects studied still varies across schools and so questions remain around the meaningfulness of some school comparisons. The Government does publish separate Progress 8 scores for English and maths and so these are perhaps more directly comparable across schools, but provide a narrower academic focus than Progress 8. Differential difficulty or generosity of subject grading can be addressed by calculating alternative Progress 8 measures where subject grade points are made statistically comparable to English and maths (which as compulsory subjects are taken by all pupils) (FFT Education Datalab, 2017). Enhancing comparability of between subject grading in this manner raises further questions, including how palatable it is to award different points to the same grade dependent on subject or whether it would help to reduce incentives for entering students for qualifications simply as 'easy' options.

However, the emphasis on EBacc subjects in the school accountability system has also raised concerns over equality of access and effects on other subjects (Parameshwaran & Thomson, 2015; Taylor, 2011). Allen and Thomson (2016) highlight the disadvantage gap in entry rates to EBacc subjects. Schools serving more disadvantaged students may find it harder to fulfil targets associated with the EBacc and relatedly in scoring highly in Progress 8. Certain schools, for instance those in more deprived areas or facing more funding pressure, may also struggle to recruit and retain the best teachers for EBacc subjects such as languages (Long & Bolton, 2017; Armitage & Lau, 2018). Although research has suggested there could be intrinsic value to studying the EBacc subjects themselves (Armitage & Lau, 2018), there is concern over whether the subject set is too restrictive or suitable for all students and schools (e.g., University Technology Colleges which focus on technical and vocational qualifications for future careers). For example, those working in educational establishments cite the pressures of Progress 8, the EBacc, and financial issues as contributing to decreasing entries to arts subjects (Johnes, 2017).

**Non-academic subjects and outcomes**

Progress 8 provides an academic summary of school performance. However, schools also influence many important non-academic student outcomes, including attitudes, behaviours, and mental health. These dimensions of school performance should be captured through the Office for Standards in Education, Children’s Services and Skills (Ofsted), the second pillar of the accountability system. However, Ofsted have often been criticised for over-relying on school academic performance data when judging schools (FFT Education Datalab, 2015a). If data on a range of the most important non-academic student outcomes were to be collected to complement academic outcomes (beyond the attendance, exclusions and destination data provided currently), this might help assist in informing more balanced judgements of schools (Clarke, 2021; Prior et al., forthcoming).

## 4. Adjustments for pupil prior attainment and pupil background

### Differing subject coverage in input and output measures

As a value-added measure, Progress 8 requires a measure of prior attainment to calculate the academic progress students make over time; attainment at KS2 fulfils this purpose. However, the pool of subjects covered by KS2 and Attainment 8, the input and outcome measures of Progress 8, are different. Specifically, whereas Attainment 8 includes student performance in a wide range of subjects as outlined above, KS2 scores are based solely upon performance in English (reading) and mathematics. This raises a conceptual challenge for Progress 8: What construct of learning are we actually measuring by Progress 8 when we are in effect comparing pupil performance in quite different subject mixes at the start and end of secondary schooling?

### Measurement error in pupil prior attainment

As with any attainment measure, KS2 test scores will contain measurement error. The bias introduced into value-added measures by classical measurement error in prior attainment operates as an overestimation of progress for students with higher true prior attainment and vice versa (Kane, 2017). School clustering of pupils with higher or lower true prior attainment will thus result in corresponding positive or negative influences on school value-added scores (Kane, 2017).

Perry (2019) suggested that attenuation bias in the national Attainment 8 relationship with KS2 introduced through KS2 score measurement error could be enough to explain the generally high performance of grammar schools compared with comprehensives. FFT

Education Datalab (2016) also cited the ‘noisy’ nature of KS2 attainment as contributing to the higher performance of grammar schools for lower KS2 attaining pupils. Achievement in the 11-plus test (determining entrance to a selective school) does not always align with higher KS2 scores, so the progress made by students with lower KS2 attainment who nonetheless pass their 11-plus is exaggerated (as their true KS2 score is above their observed KS2 score).

Cronbach’s alpha values (a measure of internal consistency or reliability) for the KS2 tests for 2018 are around 0.90 (STA, 2018). The reliability of the pupil KS2 score used in Progress 8 would presumably be somewhat higher as it is the average of pupils’ reading and maths scores. While these statistics would be viewed positively by conventional standards, given the arguments made by Perry (2019) and FFT Education Datalab (2016) further research is required to establish just how important or not measurement error is when calculating school Progress 8 scores. This could be done via formulating Progress 8 as a linear regression model of Attainment 8 and KS2 score (see Supplemental Materials) and specifying the reliability of the observed KS2 score, or by regressing Attainment 8 on latent true prior attainment relating this to the item-level KS2 data via a measurement model.

**School differences in pupil background and earlier prior attainment**

Test scores at KS2 are the only input measure used to calculate Progress 8. This was a deliberate decision; it was stipulated that other factors known to affect pupil performance were to be ignored in the model (Burgess & Thomson, 2013). In contrast, the Labour government published a school progress measure adjusting for pupil sociodemographic characteristics between 2006-2010: Contextual Value-Added (CVA) (Ray et al., 2009).

In the academic literature, adjusting for pupil prior attainment is widely considered the minimum requirement for school performance comparisons (Goldstein, 1997; Raudenbush & Willms, 1995). Most authors additionally adjust for a range of student demographic and socioeconomic background characteristics as these factors also predict pupil outcome attainment and vary between schools (Goldstein et al., 2000; Muñoz-Chereau & Thomas, 2016; Timmermans & Thomas, 2015; Wilson & Piebalga, 2008). By accounting for these characteristics, proponents argue that one moves closer to measuring the actual effects of schools on student learning. Thus, a consistent critique of Progress 8 is that even though it adjusts for prior attainment it still punishes and rewards schools for serving educationally disadvantaged or advantaged intakes as other pupil characteristics related to performance are not considered (Leckie & Goldstein, 2019; Perry, 2016; FFT Education Datalab, 2018b).

Leckie and Goldstein (2019) show that adjusting Progress 8 for student background (age, gender, ethnicity, language, special educational needs, eligibility for free school meals (FSM), deprivation) can have a marked impact on the performance of schools: a third of schools in 2015/16 would change Progress 8 bandings were the government to replace Progress 8 with a pupil background adjusted Progress 8 measure. Figure 1 (left) presents a scatter plot of schools' pupil background adjusted and unadjusted Progress 8 scores (top) and ranks (bottom) based on the 2018/19 data. While these are strong relationships ($r = 0.893$; $\rho = 0.886$), school performance nonetheless differs depending on which progress measure schools are judged by. This is shown by the substantial number of schools located away from the 45-degree line, especially in the bottom plot. Indeed, changing from unadjusted Progress 8 to background adjusted Progress 8 would lead 681 schools (21% of all schools in the country) to move up or down the national league table by 500 or more ranks, with 128 schools (4%) moving over 1,000 ranks. Bearing in mind that there are only around 3,000 secondary schools nationally, these changes are very large indeed.

One pupil background characteristic that Leckie and Goldstein (2019) did not consider was pupil KS1 score. KS1 scores might also predict pupil progress during secondary schooling: pupils with the same KS2 scores may nevertheless show different Attainment 8 scores if they differ in their KS1 scores. For example, pupils who progress rapidly from KS1 to KS2 could be more likely to continue progressing rapidly during secondary schooling compared to pupils who coasted to the same KS2 performance. If so, schools might then be argued to be unfairly advantaged if they have disproportionate numbers of pupils who made rapid progress from KS1 to KS2. Figure 1 (right) presents a scatter plot of Progress 8 scores jointly adjusted for KS1 and background (top) and ranks (bottom) against unadjusted Progress 8 scores. The relationships are only marginally weaker than when we ignore KS1 scores ($r = 0.885$; $\rho = 0.877$). Thus, in contrast to accounting for pupil sociodemographics, there appears to be little additional benefit to accounting for pupil KS1 scores when measuring secondary school performance.

While adjusting for KS1 has little effect on Progress 8 over and above adjusting for pupil sociodemographics, this will not be the case for other omitted pupil background characteristics. For instance, Dearden et al. (2011) highlighted how a lack of data on mother's education led the old CVA measure to be biased towards schools with greater proportions of mothers with higher qualifications, and this was having already controlled for several important socioeconomic characteristics. As a second example, children from advantaged socio-economic backgrounds are more likely to receive both extra support at home and sometimes private tuition outside of school. It follows that schools with a large proportion of socio-economically advantaged pupils will unfairly see the benefit of this extra investment made by parents in their Progress 8 scores. Adjusting for pupil socio-economic backgrounds as proxied by FSM would only go some way to addressing this. Richer histories of FSM eligibility could be included in a contextualised Progress 8 measure, along with other variables such as whether the pupil moved into the school recently (Thomson, 2021). However, more generally, there will always remain factors that feed into pupil performance that are unmeasured in available data sources or are themselves unquantifiable (FFT Education Datalab, 2019). This raises the question of whether we can every truly measure school effectiveness.

**Ignores intersectionality**

The current approach to Progress 8 also ignores potential interactions between pupil characteristics, such as that identified for economically disadvantaged White British pupils, who perform disproportionately worse than their advantaged counterparts (Shaw, et al., 2017; UK Parliament, 2020). Schools serving large proportions of disadvantaged White British pupils will likely be especially penalised by Progress 8. This might go some way to explaining regional patterns of low performance shown for schools in the North East and coastal schools (Leckie & Goldstein, 2019; FFT Education Datalab, 2015b). However, it is important to note that part of the heightened disparity between advantaged and disadvantaged White British pupils versus disparities for other ethnic groups arises because the average income of White British advantaged families exceeds that of other ethnic groups (Shaw, et al., 2016). Thus, this finding in part reflects a more general measurement issue surrounding the binary nature of FSM as a proxy for family economic circumstances (Jerrim, 2020). Further research is needed to explore the importance or not of accounting for intersectionality effects in Progress 8, especially as such effects were accounted for in the previously published CVA measure (Ray et al., 2009). The potential inclusion of interactions also raises further statistical considerations, such as how key predictors and their relationships with the outcome measure are to be modelled and most realistically captured (Thomson, 2021).

**Comparing the incomparable**

By adjusting for KS2 scores, Progress 8 benefits from allowing users to compare any pair of schools despite their differing student intakes. However, there are limits to the sense in making such comparisons for schools that have fundamentally different intakes and therefore little or no overlap in their KS2 score distributions. One solution would be to add a "similarity" indicator which signals the extent to which the student intakes of any schools being compared overlap or not. A warning flag might be issued to users when the schools being compared overlap very little. Related to this, the FFT Education Datalab (2021) provide a website for making bespoke school comparisons, where you can apply multiple filters across school and pupil characteristics in order to create comparisons between schools according to similarity across the chosen factors (https://schoolslikeyours.ffteducationdatalab.org.uk/). There are many innovative ideas that could potentially be applied to the Government presentation of Progress 8.

**Arguments against adjustment**

The DfE policy position against adjustment for student background characteristics (Burgess & Thomson, 2013; Department for Education, 2017) rests upon the argument that it would entrench low aspirations for disadvantaged pupils (Department for Education, 2010; Leckie & Goldstein, 2017). Such entrenchment would arise if schools set lower target GCSE performance for disadvantaged pupils compared to advantaged pupils who start with the same KS2 scores (Selfridge, 2018). It is not clear to what extent this takes place, but to the extent that it might highlights the perverse incentives that are often generated by high-stakes performance measures (Foley & Goldstein, 2012). Additionally, accounting for student background characteristics in school value-added measures could serve to remove some genuine school differences, depending on the factors which are ultimately contributing to the observed differentials. For example, if the association between poverty and lower performance chiefly reflects the impact of poorer teaching standards in deprived areas (rather than being driven by factors external to the school environment) then controlling for student disadvantage might remove this variation despite it being an important component to how well a school is serving its student population (Perry, 2016). Marks (2020) also argues that adjusting for student background characteristics may well introduce measurement issues.

## 5. Schools and pupils not covered by the analysis

### Pupils with missing scores are excluded

Progress 8 should ideally capture the influence of schools on all their pupils. However, pupils with missing Attainment 8 or KS2 scores, for instance due to pupils moving outside of English mainstream schools, are ignored. Under Progress 8 schools are not held accountable for their influence on these student's outcomes. While the median percentage of KS4 pupils included in Progress 8 for 2018/19 is 97%, with half of schools managing coverage between 93% and 98%, there is nonetheless a long tail of schools with lower coverage with some 111 schools with less than 80% coverage. These schools are not evenly distributed around the country, but rather clustered in areas with many immigrant families. While the DfE suppresses scores for schools with less than 50% coverage, it is debatable whether this is a high enough cut off. Especially as these coverage rates relate to the percentage of pupils included who were still present at the end of KS4. These percentages would be lower if they were based on all pupils who spent any time in these schools (i.e., pupils who left before the end of KS4).

### Progress 8 ignores pupil mobility

The calculation of Progress 8 is based on school membership at the end of KS4 (Department for Education, 2020a). Thus, a school is not held accountable for the performance of pupils who attend the first four years of secondary schooling, but then change to another school. In contrast, where a pupil moves into their school for only the final year, they are held accountable for the entirety of the progress that pupil has made over all five years of secondary schooling. Progress 8 therefore ignores pupil mobility. Whilst we would expect most pupils to attend one school for the duration of secondary schooling (Leckie, 2009), the proportion of mobile pupils is likely to vary between schools. A concern with ignoring pupil mobility is that this may be incentivising off-rolling, a gaming practice removing students from a school through unofficial channels to improve scores on performance metrics. Off-rolling has been increasing in recent years (Hutchinson & Crenna-Jennings, 2019; Guardian, 2020a).

Leckie (2009) uses multilevel modelling to show that the rank order of school value-added scores is sensitive to whether a relatively mild degree of pupil mobility was accounted for or not. The contribution of pupil data to the school effects were weighted by the number of years pupils spent in each of their schools. Analysis by FFT Education Datalab (2018c) explored a similar idea to reweight Progress 8 scores. Their analysis, based on the cohort completing GCSEs in 2017, showed that accounting for mobility in this way reduced the advantage seen on average in London schools (FFT Education Datalab, 2018d).

**Relevance of Progress 8 to school types serving different age ranges**

The calculation of Progress 8 based on school membership at the end of KS4 also has consequences for its relevance as a measure for different schools. Progress 8 disadvantages schools with non-standard age-ranges, holding them to the same standard as schools teaching throughout secondary schooling, despite having less time with students in which to influence performance. This is therefore an important consideration when comparing Progress 8 across school types (Figure 2): the average Progress 8 scores for Studio schools, University

Technical Colleges (UTCs), and especially colleges of further education are all notably lower than that shown by other school types. These schools typically only teach pupils from age 14, taking pupils that have often struggled at school up to this point. The emphasis in these types of schools is also more focused on vocational education in preparation for future careers rather than the academic subjects prescribed by Attainment 8. The DfE do note on their school performance website that other measures such as pupil destinations are more appropriate for these schools, but their scores are nonetheless still published and thus likely compared to others by many users.

In 2018, the DfE took steps to recognise that the blanket application of Progress 8 to schools serving non-standard age ranges may be unfair. The ‘floor standard’, a threshold score below which schools are judged to be underperforming and may come under extra scrutiny from the school inspectorate, was to no longer apply to UTCs, Studio schools as well as Further Education colleges with age 14-16 provision. From September 2019, the floor standard was then dropped for all schools (Department for Education, 2020a). The removal of this deterministic link between Progress 8 and the inspectorate system, lessens the stakes attached to this measure. This may in time alleviate some of the likely negative side effects of Progress 8, such as teaching to the test, pupil and teacher stress and gaming behaviour, for which high stakes accountability systems are frequently criticised (Amrein-Beardsley, 2014; Foley & Goldstein, 2012; Koretz, 2017; NAHT, 2018).

## 6. Presentation, choice of statistical model, and calculation of statistical uncertainty

### Schools only account for around a tenth of the variation in pupil progress

An important point missing from the presentation of school Progress 8 scores is any statement about the overall importance of schools as a driver of the national variation in pupil progress. School Progress 8 scores measure the mean pupil progress in each school.

However, school mean differences in 2018/19 only accounted for 12% of the total variation in pupil Progress 8 scores, with the remaining 88% occurring within rather than between schools. Another way to understand this point is to think about the extent to which the national average Attainment 8 score would increase if all schools with negative Progress 8 scores (half of all schools) were improved to have scores equal to the national average school. The national average Attainment 8 score would only increase by 1.75 points from 47.75 to 49.50 equivalent to a pupil scoring just one grade higher in two out of their eight subjects. This is not to say that schools are unimportant, but that they are only part of the explanation for why some pupils progress more rapidly than others. If this is not well understood, there is risk that too much attention is placed on comparing school Progress 8 scores when much of the potential for increasing pupil progress lies within rather than between schools.

**Scores are not readily understandable**

How should we interpret a school with a Progress 8 score equal to 0.53? What are the units of measurement? Is this a big score compared to other schools? The magnitude of a Progress 8 score is not readily understandable, making these questions hard to answer. One has to consult the technical documentation to understand that pupils in this school score, on average, just over half a grade higher per Attainment 8 subject than other pupils nationally with similar prior attainment (Department for Education, 2016). However, even if users realise this meaning, there is no clear guidance as to whether this is a large effect either in absolute or relative terms, compared with other schools. More needs to be done.

Expressed as an effect size, a Progress 8 score of 0.53 corresponds to a mean difference of 0.27 SD in pupil Attainment 8 (the national SD of pupil Attainment 8 scores in 2018/19 was 1.93 SD). Thus, pupils in our example school on average score 0.27 SD higher at GCSE than pupils with the same KS2 scores attending the average school ($= 0.53/1.93$).

While, perhaps helpful to researchers, communicating Progress 8 scores in this way is unlikely to be accessible to users. One response would be to express Progress 8 scores in terms of GCSE grades. In terms of our example school, a Progress 8 score of 0.53 approximately corresponds to a pupil who might score eight 4's in the national average school now scoring four 4's and four 5's in this high progress school (where a grade 4 and grade 5 in English and maths are double weighted).

Another possibility would be to apply the Education Endowment Foundation's (EEF's) additional months' progress measure for evaluating school-based interventions (Education Endowment Foundation, 2021). This measure assumes that the average annual increase in student attainment at a given age is approximately equal to 1.0 SD difference in attainment at that age. An effect size of 0.27 SD therefore translates to pupils being 0.27 of a year or three months ahead of those in the average school and EEF describe this as a 'moderate impact' ($= 12 \times 0.27/1.00$). The EEF (2018) note, however, that the rate that the average annual increase in student attainment decreases as children grow older and that a value of 0.4 SD is perhaps more reasonable than 1.0 SD by age 14 in which case an effect size of 0.27 SD would translate to pupils being a much more impressive eight months ahead of those in the average school ($= 12 \times 0.27/0.40$). Figure 3 plots the national distribution of Progress 8 scores expressed as months' progress relative to the average school separately based on these alternative assumptions regarding the average annual increase in student attainment. The choice of assumption is clearly of fundamental importance.

The relative position of schools within the national distribution (percentile ranks) could also be communicated to users. A Progress 8 score of 0.53 corresponds to the 87th percentile: pupils in this school make, on average, more progress than 87% of schools in England. However, this solution comes with its own problem. A fixed difference in percentile ranks corresponds to much larger Progress 8 score differences in the tails of the national distribution than in the middle of the distribution. For example, while pupils in a school at the 95th percentile score on average 0.32 grades per subject higher than equivalent pupils in a school at the 85th percentile, pupils in a school at the 55th percentile only scores 0.11 grades per subject higher than equivalent pupils in a school at the 45th percentile.

Finally, and most importantly, even if the magnitude of Progress 8 scores can be made readily understandable, they are still simple one-number summary measures which do not shed any light as to which school policies and practices lead schools to score the way they do. Progress 8 is fundamentally a 'black box'.

**95% confidence intervals are not readily understandable**

The 95% confidence intervals provided with Progress 8 scores are also challenging for users to interpret and have often been discarded in media presentations of the data. The presentation of Progress 8 attempts to address this by displaying bandings of school performance which are a joint function of the magnitude of the score and its confidence interval (as detailed in Section 2). Only schools which are statistically significant can appear in the more extreme bandings. Therefore, this approach attempts to factor statistical uncertainty into the decisions that users make, with evidence suggesting parents do focus on the bandings when making school choices (Menzies & Jerrim, 2020).

A second issue relates to the problem of multiple comparisons, that is, for accountability purposes, some 3,000 schools' Progress 8 scores are compared to the average score of 0. Thus, even if there were no differences in the unobserved 'true' performance of schools, then we would still expect to see 95% confidence intervals suggesting that 1 in 20 schools (approximately 150) are performing significantly differently above or below average (Type I errors: 'false positive' results). Given the high stakes associated with such statements, it may be prudent to therefore widen the 95% confidence intervals to counteract falsely singling out individual schools in this way. The Bonferroni correction is one of several methods that could be explored to do this. The trade off, of course, is that in doing so we simultaneously reduce the power to detect schools whose unobserved ‘true’ performances truly do differ from average (Type II errors: false ‘negative’ results).

A third issue is that the presented confidence intervals, while correct for comparing a single school to the overall average, are not correct for comparing a single pair of schools to one another. That is, two overlapping 95% confidence intervals do not automatically imply the performance of the two schools are not statistically significantly different: a small degree of overlap can be tolerated. Goldstein & Healy (1995) therefore propose, for the purpose of making a single pairwise comparison, an adjustment to narrow the 95% confidence intervals so that the scores of two schools whose confidence intervals are just touching are now borderline significantly different. More generally, however, many users will again want to make multiple pairwise comparisons, and so here too further adjustments may again be required to counteract the problem of multiple comparisons.

An alternative to simply stating or plotting the 95% confidence intervals is graphical presentation of Progress 8 scores via a funnel plot. This would depict visually the amount of variation in Progress 8 scores expected by chance as a decreasing function of school size, helping flag schools for further investigation whose scores vary beyond this (FFT Education Datalab, 2015c). Another approach, particularly relevant when making bespoke comparisons between small sets of schools’ Progress 8 scores and 95% confidence intervals, would be to translate this information into probabilistic statements as to how statistically certain any difference in measured progress was between any pair of schools (Leckie and Goldstein, 2011).

**Schools make different progress with different pupil groups**

The DfE also reports Progress 8 by pupil sub-groups, for instance by prior attainer (low, middle, high), disadvantage, English as an additional language, and gender. This reporting recognises that schools may be differentially effective for different pupil groups. However, the number of pupils in these groups within many schools is often low and so the resulting scores will be less reliable than overall scores. This can be seen in Figure 4, which shows the ranked school Progress 8 scores and 95% confidence intervals in 2019 for all, disadvantaged, and non-disadvantaged pupils. All three plots show that many schools cannot be statistically distinguished from the overall average. However, the confidence intervals are notably longer for the disadvantaged subgroup, reflecting the smaller number of pupils per school in this group. Statistical disclosure prevents presentation of further breakdowns by ethnic groups and other pupil characteristics where the number of pupils in many categories is low. A subtle further issue with reporting Progress 8 separately for low, middle, and high prior attainers is that the mean prior attainment of pupils in each of these groups itself varies across schools limiting the meaningfulness of such comparisons: we are not comparing like with like.

**Choice of statistical model**

In the academic literature, the most common approach to studying school effects on pupil attainment is via estimating multilevel linear regressions of pupil outcome attainment on prior attainment including a school random intercept effect to predict school performances (Aitkin & Longford, 1986; Goldstein, 1997; OECD, 2008; Reynolds, et al., 2014). In contrast, the approach underlying Progress 8 is equivalent to fitting a conventional ‘single-level’ linear regression of pupil outcome attainment on prior attainment, then averaging the predicted pupil residuals to the school level (see Supplementary Materials). This approach is favoured by the DfE because it is simpler to explain to users and because it results in zero mean pupil Progress 8 scores at every level of prior attainment (Burgess & Thomson, 2013).

Zero mean pupil Progress 8 scores at every level of prior attainment is on first inspection intuitively appealing as then a pupil Progress 8 score of 1.0 for a low prior attainer has the same meaning as it does for a high prior attainer, namely that both pupils score 1.0 grade higher per subject than the average pupil nationally who started with the same prior attainment. In contrast, the multilevel linear regression approach shows negative mean pupil Progress 8 scores at low levels of prior attainment and positive mean pupil Progress 8 scores at high levels of prior attainment (Figure 6). This divergence of results between the two regression models indicates that school mean prior attainment is positively correlated with the school effects. This is consistent with the operation of two potential social processes: positive peer effects associated with being educated among higher prior attaining peers; and higher prior attaining pupils selecting into more effective schools, or equally more effective teachers selecting into schools with higher prior attaining peers (Castellano et al., 2014). For school accountability purposes it would seem desirable to estimate school effects having controlled for these two social processes. However, neither approach does so, leading the true variability in school effects to be underestimated, but this will prove especially the case for the single-level approach employed by Progress 8. Indeed, Figure 7 shows that shifting from a linear regression model to a multilevel linear regression model would lead school Progress 8 scores for schools with higher mean prior attaining intakes, most notably Grammar schools, to increase by a non-trivial 0.1-0.2 grades per subject. However, given just how little overlap there is in the pupil prior attainment distribution between Grammar and other schools, it is questionable how meaningful it to compare these two school types in this way irrespective of whether a single-level or multilevel linear regression is employed.

Putting aside arguments about peer and selection effects, two often stated benefits of the multilevel modelling approach are that it is more extendable to the study of differential school effects (Nuttall, et al., 1989; Strand, 2016), and that the estimated school effects are ‘shrinkage’ (empirical Bayes) estimates that pull the effects for small schools towards the overall average and thus deter overinterpretation of otherwise erratic results (Goldstein, 1997). We focus here on illustrating this second point, noting however that shrinkage can also be applied without fitting a multilevel model (Clarke, 2021). Specifically, we apply shrinkage separately to the school Progress 8 scores that we plotted in Figure 4 separately for all, disadvantaged, and non-disadvantaged pupils. Figure 5 presents plots of the difference between these shrunken and unshrunken Progress 8 scores against school size. The figure shows the expected pattern by which small schools scores are shrunk considerably more than big schools. This pattern is particularly pronounced for scores for disadvantaged pupils once again highlighting how the Progress 8 summaries for these pupils are unreliable estimates of their true performance and therefore questioning the reporting of statistics based on such small numbers of pupils.

**Schools may influence the variance in pupil progress**

Progress 8 reports the average pupil progress in each school. This is the predominant approach in the academic literature but one which ignores other potential aspects of a school’s influence. For example, schools may also influence the variance in pupil progress (Leckie et al., 2021). Thus, two schools which may appear equal in terms of average pupil progress may appear quite different when one explores the variability in progress about these averages. In one school, the variation in pupil progress around the school average may be very low suggesting that the school is educating their pupils in a consistent way. In the second school, the variation may be very high suggesting that the school is educating their pupils in a more erratic fashion. One way to address this issue would be to additionally report the percentage of pupils in each school with very low and very high Progress 8 scores.

## 7. Instability of school effects

### Scores are unstable over time

Research on school effects suggests they are unstable over time, adding uncertainty to the identification of consistently effective or ineffective schools (Leckie, 2018; Marks, 2015; Thomas, et al., 1997). This is also the case for Progress 8 where the correlation between scores in 2016 and 2019 is just 0.69 (Table 2). Changes in Progress 8 scores will reflect not just the improving or worsening effectiveness of the school, but also changing pupil intake characteristics, sampling variability, marking unreliability, and external changes to the accountability system (Menzies & Jerrim, 2020). Reliance on a single year of data amplifies the consequences of this instability and increases the pressure faced by schools and pupils. Small schools' effects are likely to be particularly unstable and may show large changes year on year. Figure 6 demonstrates the wider spread for smaller schools when examining the change between the 2018 and 2019 Progress 8 scores. Additionally, given the usual presentation of scores in rank order by the media in simplistic league tables, which may give little prominence to statistical uncertainty, estimates for small schools may be at particular risk of overinterpretation as they tend to appear towards the extremes of these tables.

### Past performance of schools a limited guide to their future performance

School performance measures, including Progress 8, are based on the performance of schools for the most recent cohort of students completing their secondary education, whereas the information relevant for school choice is how well schools will perform some seven years into the future when the children choosing complete their secondary schooling (Leckie and Goldstein, 2009). Table 2 shows the diminishing association between Progress 8 scores over time. The correlation between the 2019 and 2018 scores is 0.87 compared with 0.69 between 2019 and 2016. The correlation relevant for school choice would be expected to be substantially lower still (we cannot show this here, due to Progress 8 scores only currently being available for a four-year time horizon, whereas a seven-year time horizon is needed). The instability of Progress 8 over time, combined with this long time lag, means the future performance of schools is far less certain than their current performance (Allen & Burgess, 2011; Leckie & Goldstein, 2009). The Government, however, provide few, if any, warnings about this issue. One response would be to statistically dampen the scores and widen the 95% confidence intervals to reflect this increased uncertainty (Leckie and Goldstein, 2009).

**Multi-year averages may help smooth noisy performances**

Multi-year averages potentially provide a way to address some of the concerns around the instability of school effects over time, especially relating to sampling variability. Averaging across years would enhance the signal of school performance by smoothing out noise year to year. The resulting average will also have a narrower confidence interval. The public can already access the Progress 8 scores for schools for the previous two years alongside the current year on the DfE's school performance website, however, no overall summary is presented. Meta-analysis provides a way to average this published data without reanalysing the underlying pupil level data (Jerrim, 2019). To demonstrate the technique, Figure 8 presents the Progress 8 scores for two schools in Bristol between 2017 and 2019, plus an estimate for the overall average progress of these school across the three years. For St Bede's Catholic College (top plot) the confidence interval for the overall estimate (the width of the diamond) is smaller than for any individual year: we are more confident in judging the school as performing above average based on three years of data than on one. Indeed, judgement based on either of the last two years would rank the school as not statistically distinguishable from average. For Bristol Free School (bottom plot), Progress 8 scores are very volatile from one year to the next due to the small school size. Pooling information across the three years results in arguably a more meaningful summary of performance than focusing on any year in isolation.

## 8. Implications of COVID-19

### No 2020 or 2021 school performance tables

The disruption to schools and examinations caused by the COVID-19 pandemic raises further challenges and questions for Progress 8. The cancellation of 2020 and 2021 GCSE exams mean that Progress 8 will not be published for two years. Given the furore around the 2020 teacher moderated grades (The Guardian, 2020b; BBC News, 2020b) and the concerns around differential lost learning for 2021 students (Department for Education, 2020e) this seems an entirely sensible decision.

### Impact of missing scores on future Progress 8 school performance tables

While we note that a consultation into the future of GCSEs is underway (Schools Week, 2021), the assumption is that GCSE exams in some form and therefore Progress 8 or a similar accountability measure will resume from 2022 onwards. However, the cancelation of 2020 and 2021 KS2 tests will also prevent Progress 8 from being published in its current format in 2025 and 2026; the necessary prior attainment measure will not be available. While 2022 KS2 tests are scheduled to run, question marks will hang in the air as to their accuracy given differential lost learning across schools in 2020/21, thus potentially impacting upon Progress 8 scores in 2027. In 2010, there was a partial boycott of KS2 tests, with around a quarter of schools refusing to participate (BBC News, 2010). Teacher assessments were used in place of test results for schools participating in the boycott. However, the use of teacher assessments in lieu of tests is problematic as the assessments would likely be biased upwards relative to the unknown test scores; analysis of the 2010 teacher assessments highlighted that a higher percentage of pupils in boycotting schools received the top grade compared with those taking the standard tests (Schools Week, 2015). Schools teaching large proportions of pupils affected by the boycott therefore saw a dip in their 2015 school value-added scores. If a Progress 8 measure is to be published in 2025 and 2026, other options would need to be considered. For any potential solution, the resulting Progress 8 scores are likely to be far more volatile than surrounding years.

### Differential impact of COVID-19 on pupil groups

There are also concerns that schools could be unfairly penalised due to the differential influence of lockdown on different pupil groups (Andrew, et al., 2020; Cullinane & Montacute, 2010). School closures have widened the attainment gap between disadvantaged students and their peers (Education Endowment Foundation, 2020). Therefore, differences in Progress 8 arising from differences in the composition of pupil characteristics are likely to be exaggerated in 2022, heightening the need to adjust for student background characteristics. Additionally, some schools may try to cope with the continuing disruptions of the pandemic and lockdown by entering pupils for fewer subjects in 2022 (The Times, 2020b). Attainment 8 and therefore Progress 8 will heavily penalise schools for doing this by assigning a zero score for any unfilled slots. This suggests further sources of Progress 8 volatility across years.

## 9. Discussion

In this article, we have reviewed and evaluated the Progress 8 secondary school performance measure used in England to hold schools to account and to support parental school choice. We have highlighted strengths of Progress 8 over the previous 5A*-C measure, notably: accounting for school differences in prior attainment at intake; encouraging focus on students across the ability distribution rather than just at the GCSE grade C/D boundary; and presenting statistical uncertainty not only through 95% confidence intervals but via colour-coded bandings. However, we also noted several weaknesses of Progress 8, including: the pressures on schools and pupils induced by excessive emphasis on EBacc subjects; biases introduced through the lack of sufficient contextualisation for variation in pupil backgrounds across schools; reduced accountability for mobile pupils; the lack of understandable effect size information; and the instability of Progress 8 scores over time and its consequences for parental school choice. Informed by our review, we propose six simple recommendations to improve Progress 8 and school accountability in England.

**Six recommendations to improve Progress 8**

*Recommendation 1: Present a less EBacc focussed Progress 8 measure alongside Progress 8 to present a more holistic picture of school performance relevant to more schools and pupils.* Progress 8 emphasises a strong focus on EBacc subjects. While there are arguments in favour of a focus on a narrower range of traditional academic subjects (Wolf, 2011), there are also concerns over equality of access, its impact on other subjects and whether it makes a suitable subject choice set for all pupils and schools (Section 3). Given these concerns we recommend presenting a less EBacc focussed version of Progress 8 next to the current Progress 8 measure so that, where relevant, schools would be freer to pursue more varied academic curriculums most suitable for their pupils without the current immediate negative repercussions through Progress 8. For example, this could be achieved by reducing the number of EBacc slots in Attainment 8 from three to two and increasing the number of open slots from three to four.

*Recommendation 2: Present a pupil background adjusted Progress 8 measure alongside Progress 8 to provide a picture of school performance informed by school context.*

There are a range of arguments for and against adjusting Progress 8 for student background characteristics (Leckie & Goldstein, 2019; Perry, 2016; FFT Education Datalab, 2018b) (Section 4). We therefore recommend presenting an adjusted and an unadjusted Progress 8 measure side-by-side rather than presenting just one measure or the other. Progress 8 holds schools accountable for the different progress made by various pupil groups nationally, whereas a pupil background adjusted Progress 8 measure views these as the result of broader economic, social, and cultural differences that are the responsibility of society to address. The provision of both unadjusted and adjusted scores offers a way to recognise both perspectives. Additionally, providing information for why a school's score changes depending on choice of measure would then give a fuller picture of what is happening in schools and the potential drivers of measured performance. For instance, where a school's score improves with adjustment for background characteristics, this would suggest the school is teaching a cohort of educationally disadvantaged pupils.

*Recommendation 3: Recognise pupil mobility by making school Progress 8 scores an average of all pupils who attended each school, weighted by their time in each school.*

Progress 8 ignores pupil mobility between schools, implicitly assuming all students remain in the same establishment throughout secondary schooling (Section 5). The consequence being that schools are not held accountable for all the pupils that they have taught. Following the work of the FFT Education Datalab (2018c), our third recommendation is to account for school mobility in Progress 8 by weighting a pupil's score according to the number of years spent attending that school, proportionately assigning their score to multiple schools where moves have taken place. A measure weighted in this way should provide a more accurate portrayal of school performance and may help to de-incentivise off-rolling gaming behaviour.

*Recommendation 4: Communicate more clearly the relative importance of school Progress 8 scores in explaining the overall variation in pupil progress and the magnitude of each school's individual Progress 8 score.*

The current presentation of Progress 8 does not sufficiently communicate that the variation in school Progress 8 scores accounts for only around a tenth of the overall variation in pupil progress across England. Thus, even when two schools appear quite different in their school Progress 8 scores, the progress made by pupils in these two schools will nonetheless overlap substantially. School attended is only a small explanation for why each child shows the progress they do. It is also difficult to gauge the meaning of a school's Progress 8 score. What, for example, does a score of 0.53 mean to most users? We therefore recommend that Progress 8 scores are translated into more readily understandable metrics. One way this might be done is by communicating how a pupil's GCSE grades might change if they attend different schools. Another possibility is to communicate how many months further ahead they would be if they attend different schools. Additionally, presenting percentile ranks would then communicate the position of each school in the national distribution.

*Recommendation 5: Increase warnings regarding the substantial uncertainty in using Progress 8 to predict the future performance of schools.*

Progress 8 scores are also noisy estimates of school performance. The presented 95% confidence intervals reveal the middle 40% schools are not statistically distinguishable from the overall average school. The use of colour bandings helps communicate this. However, for supporting school choice the comparisons of interest are specific pairwise comparisons, but these are not accompanied by any statements of statistical uncertainty. We recommend that 95% confidence intervals and colour bandings are also added here to protect users form overinterpreting small differences between pairs of Progress 8 scores. More fundamentally, however, the current performance of schools is only of interest in so far that it predicts the future performance of schools for when those choosing are in attendance (Section 7). The instability of Progress 8 scores over time means that there is even more statistical uncertainty in making such future predictions. We therefore recommend that prominent warnings are also added to the school performance website to explain this currently ignored issue.

*Recommendation 6: Report multiyear averages for Progress 8 alongside current single year summaries to illustrate and combat the instability of school performance over time.*
Multi-year averages may help to address concerns with Progress 8 relating to the instability of scores over time and the overinterpretation of small schools' scores. Basing decision making on multi-year averages may allow schools to plan more sensibly over the longer term rather than constantly needing to find quick solutions in response to bouncing Progress 8 scores. Multi-year averages are easy to implement as demonstrated in Figure 7 where we calculated three-year averages with associated 95% confidence intervals directly as a function of the published school performance data. Different numbers of data years can be accommodated, and different weighting schemes can be used. For example, more weight can be given to more recent years and where school size varies across years, this can also be weighted into the presented average.

**Broader issues surrounding school accountability in England**

Our focus in this article has been on reviewing and evaluating statistical strengths and weaknesses of Progress 8, and thus the secondary school accountability system in England. However, many of the issues we have raised are also relevant to the primary school accountability system in England. Indeed, given that Primary schools are substantially smaller than secondary schools, some of these issues will be even more important, such as the instability of performance measures over time and the inability to draw conclusions about schools with statistical confidence. Our work is also relevant for the new primary school progress measure which will use the incoming reception baseline assessments (STA, 2020) as its measure of prior attainment. For example, here concerns about the impact of potential measurement error of the prior attainment adjustment will be greater given the young age of the children involved (4/5). The potentially subjective nature of teacher judgements and whether this may favour certain pupils and schools is also a cause for concern.

Furthermore, while we believe that our proposed recommendations have the potential to address some of the statistical concerns we have raised with Progress 8, it is important to reiterate the broader long-standing concerns with the way school performance data has been used to inform school accountability in England over the last three decades. Throughout this review we have noted issues related to the high-stakes accountability system in which Progress 8 is situated, often producing deleterious consequences such as gaming behaviour and perverse incentives (Amrein-Beardsley, 2014; Foley & Goldstein, 2012; NAHT, 2018; OECD, 2008). Our recommendations might possibly help alleviate some of these effects, for instance improved communication that schools only account for a tenth of the variation in pupil progress and that therefore many differences between schools scores are qualitatively very small (recommendation 4) might help reduce the excessive focus on rankings in school performance tables. Similarly, holding all schools attended accountable for the performance of a pupil rather than only the last school attended (recommendation 3) might help alleviate perverse incentives to off-roll pupils. However, we recognize that most of these long-standing concerns will likely remain as long as the stakes attached to high Progress 8 scores remain so high.

**Tables**

Table 1.

Progress 8 school performance table results for Bristol secondary schools for the 2018/19 year.

| School name | Number of pupils at end of KS4 | Attainment 8 score | Number of pupils included in Progress 8 | Progress 8 score | Progress 8 confidence interval | Progress 8 banding |
|---|---|---|---|---|---|---|
| Redland Green School | 197 | 60.8 | 181 | 0.53 | 0.35 to 0.72 | Well above average |
| Colston's Girls' School | 140 | 56.7 | 132 | 0.51 | 0.29 to 0.73 | Well above average |
| Bristol Metropolitan Academy | 165 | 48.2 | 153 | 0.50 | 0.30 to 0.71 | Well above average |
| Bristol Cathedral Choir School | 122 | 56.4 | 115 | 0.36 | 0.12 to 0.59 | Above average |
| Fairfield High School | 146 | 46.7 | 131 | 0.35 | 0.13 to 0.57 | Above average |
| St Mary Redcliffe and Temple School | 213 | 55.0 | 206 | 0.28 | 0.11 to 0.46 | Above average |
| Bristol Free School | 143 | 52.6 | 139 | 0.27 | 0.05 to 0.48 | Above average |
| Bristol Brunel Academy | 204 | 44.8 | 190 | 0.20 | 0.02 to 0.38 | Above average |
| St Bede's Catholic College | 180 | 57.5 | 178 | 0.17 | -0.02 to 0.36 | Average |
| Orchard School Bristol | 123 | 40.4 | 111 | -0.01 | -0.25 to 0.23 | Average |
| Oasis Academy Brislington | 140 | 40.9 | 133 | -0.04 | -0.26 to 0.18 | Average |
| Bridge Learning Campus | 98 | 38.6 | 96 | -0.05 | -0.31 to 0.21 | Average |
| Cotham School | 214 | 46.0 | 201 | -0.07 | -0.25 to 0.10 | Average |
| Oasis Academy John Williams | 166 | 43.9 | 162 | -0.09 | -0.29 to 0.11 | Average |
| The City Academy Bristol | 114 | 31.4 | 81 | -0.12 | -0.40 to 0.16 | Average |
| Steiner Academy Bristol | 19 | 49.1 | 10 | -0.32 | -1.12 to 0.47 | Average |
| Bedminster Down School | 174 | 39.4 | 171 | -0.39 | -0.58 to -0.20 | Below average |
| Oasis Academy Brightstowe | 133 | 36.1 | 120 | -0.39 | -0.62 to -0.16 | Below average |
| St Bernadette Catholic Secondary School | 130 | 42.2 | 126 | -0.45 | -0.68 to -0.23 | Below average |
| Henbury School | 122 | 36.5 | 106 | -0.55 | -0.80 to -0.31 | Well below average |
| Ashton Park School | 199 | 39.0 | 196 | -0.60 | -0.78 to -0.42 | Well below average |
| Merchants' Academy | 147 | 33.6 | 145 | -0.68 | -0.89 to -0.47 | Well below average |

Source: https://www.compare-school-performance.service.gov.uk/

Note:

Progress 8 bandings are defined as follows. ‘Well above average’: Progress 8 score above 0.5 and 95% confidence interval above zero; ‘Above average’: Progress 8 score higher than 0 but lower than 0.5, and 95% confidence interval above zero; ‘Average’: 95% confidence interval includes zero; ‘Below average’: Progress 8 score higher than -0.5 but lower than 0, and 95% confidence interval below zero; ‘Well below average’: Progress 8 score lower than -0.5 and 95% confidence interval below zero.

Table 2.

Pairwise correlations between school Progress 8 scores over time. Sample sizes given in italics.

| | 2016 | 2017 | 2018 | 2019 |
|---|---|---|---|---|
| | 1.00 | | | |
| 2016 | *3098* | | | |
| | 0.77 | 1.00 | | |
| 2017 | *2973* | *3133* | | |
| | 0.72 | 0.81 | 1.00 | |
| 2018 | *2811* | *2965* | *3165* | |
| | 0.69 | 0.77 | 0.87 | 1.00 |
| 2019 | *2664* | *2812* | *3008* | *3196* |

**Figures**

Figure 1.

Scatterplots of school average adjusted and unadjusted Progress 8 scores (first row) and ranks (second row) with Pearson and Spearman rank correlations.

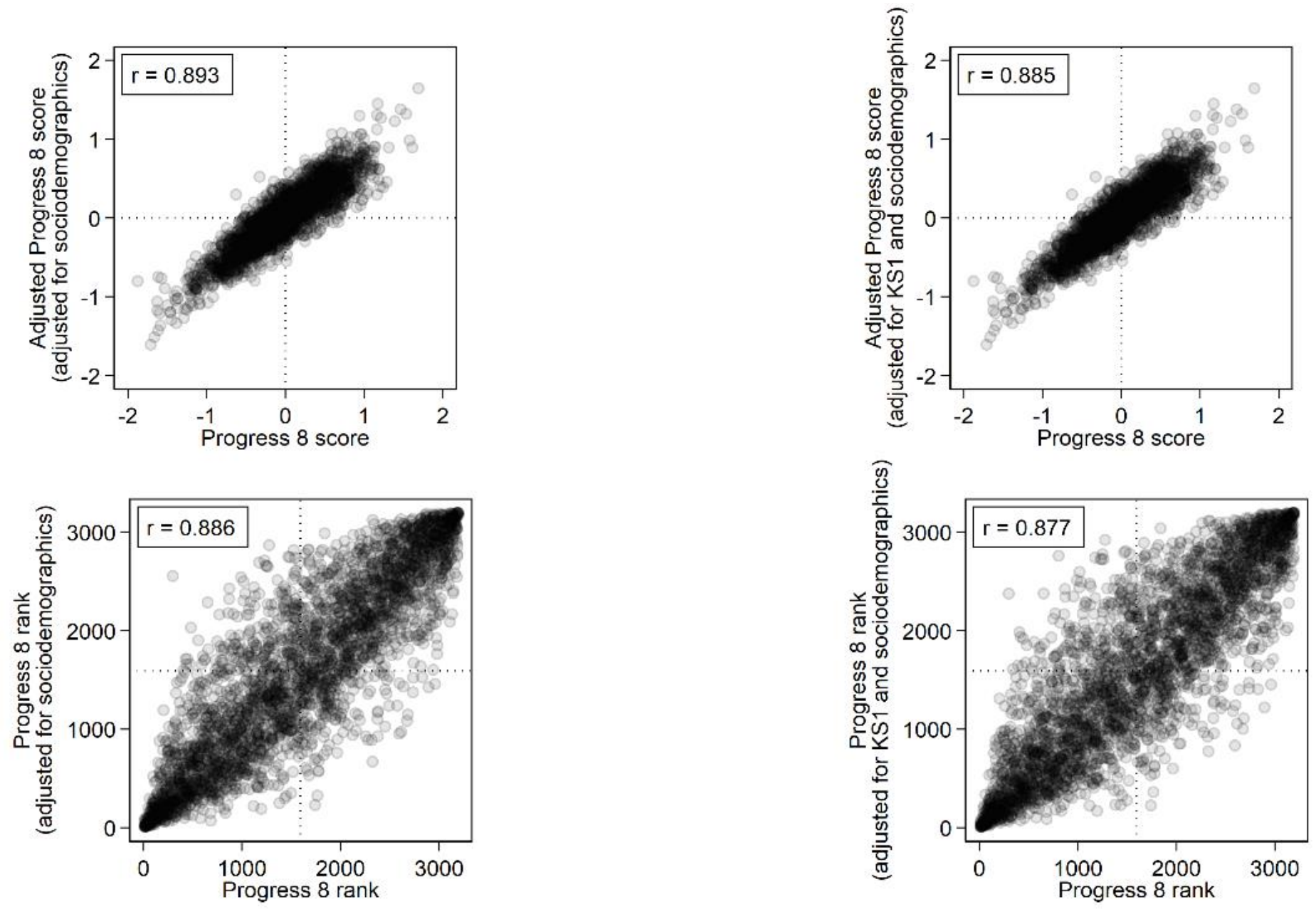

Figure 2.

Average Progress 8 scores by school types for mainstream schools in 2019.

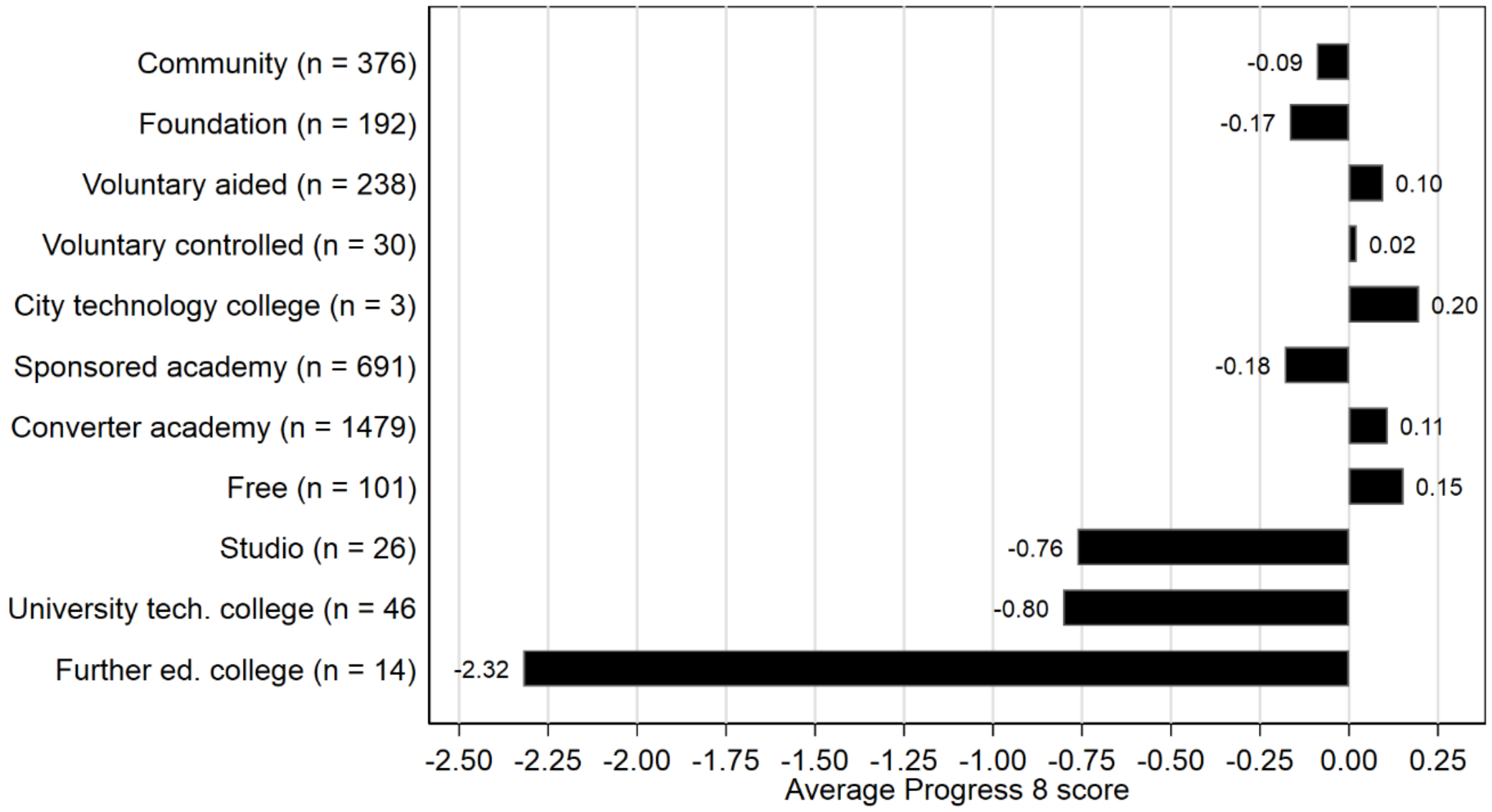

Figure 3.

Distribution of Progress 8 scores expressed using the EEF months' progress measure relative to the average school.

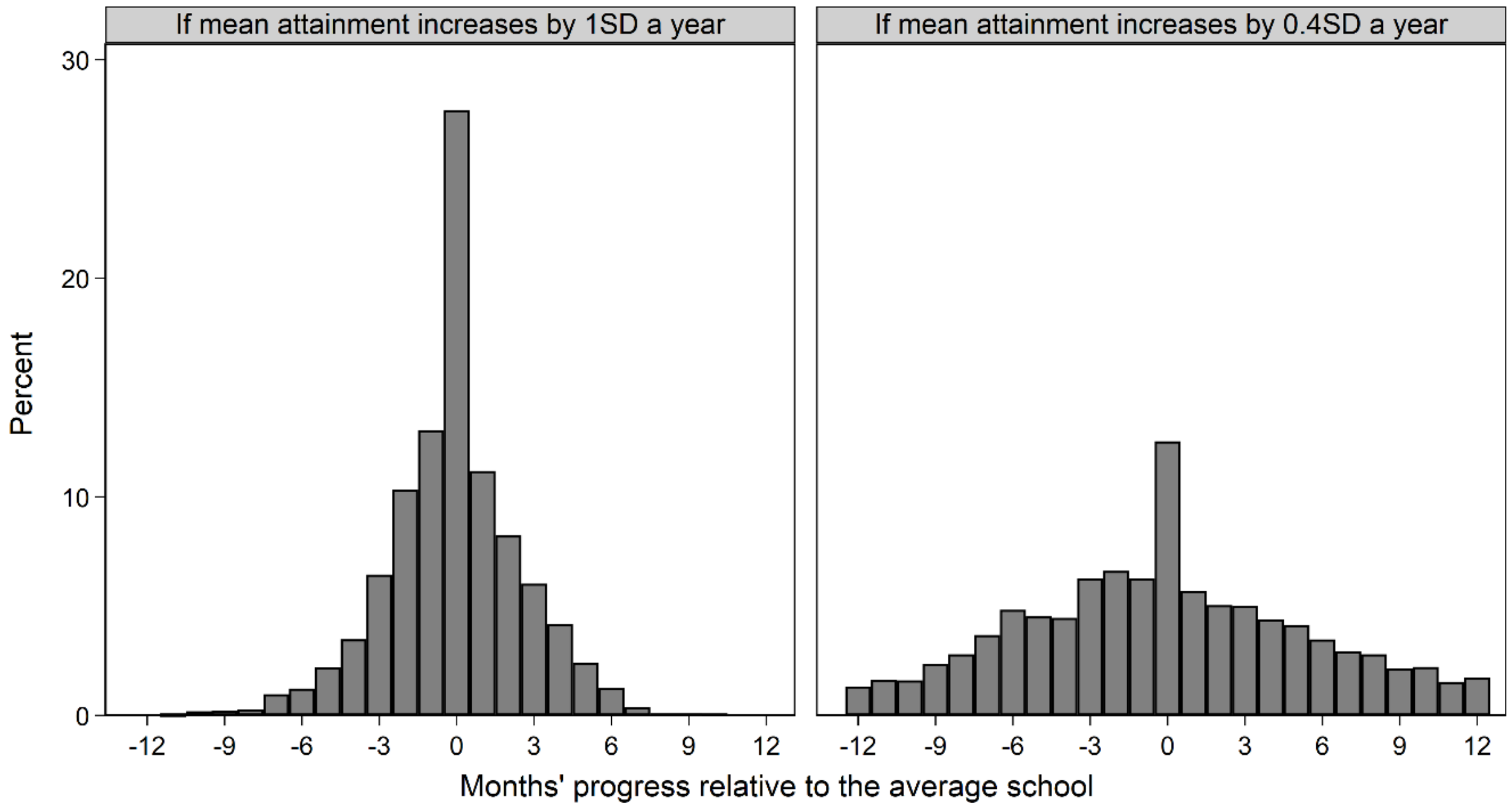

Figure 4.

Ranked school Progress 8 scores with 95% confidence intervals in 2019 for all, disadvantaged, and non-disadvantaged pupils.

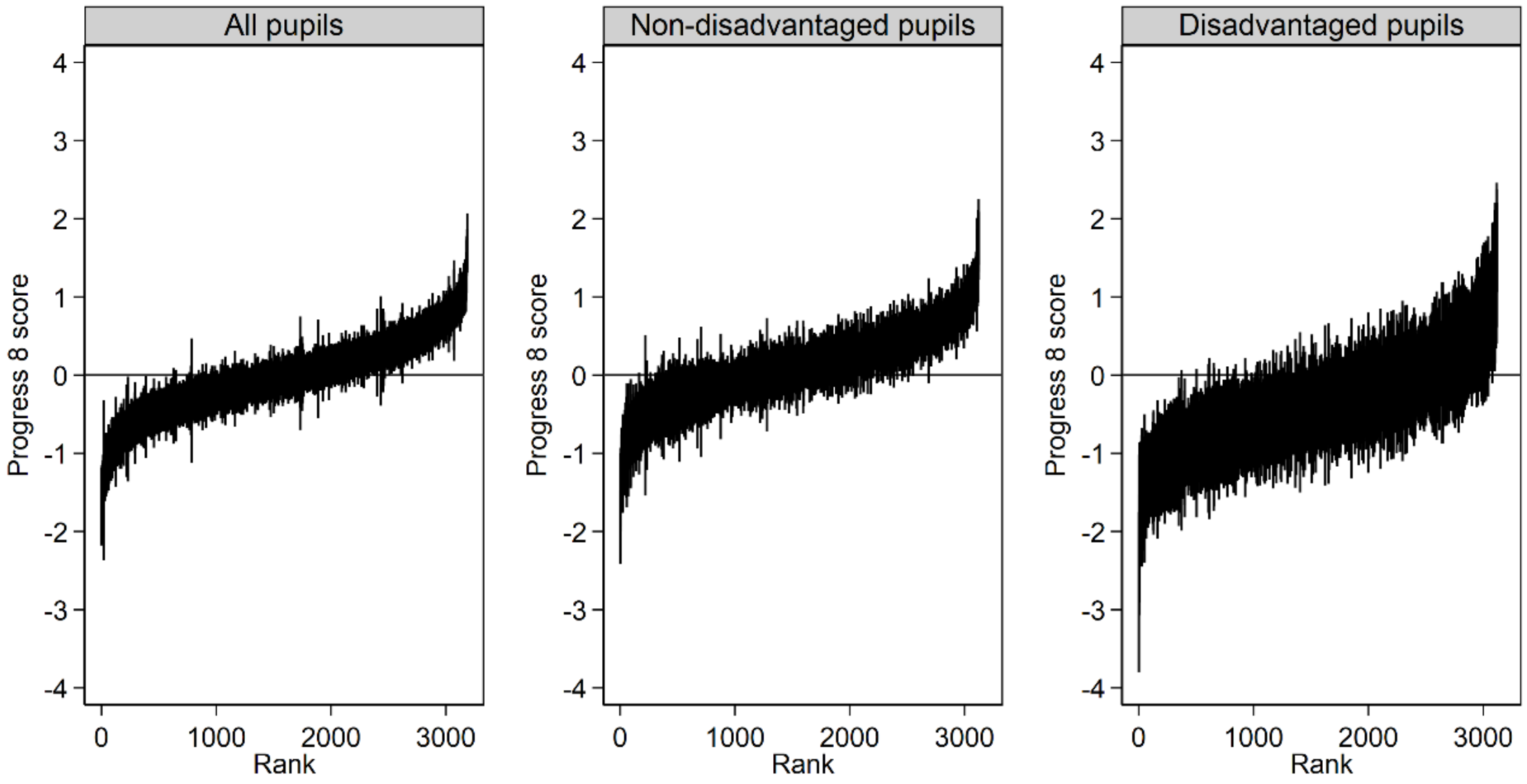

Figure 5.

Scatterplot of the difference between shrunken and unshrunken Progress 8 scores against school size.

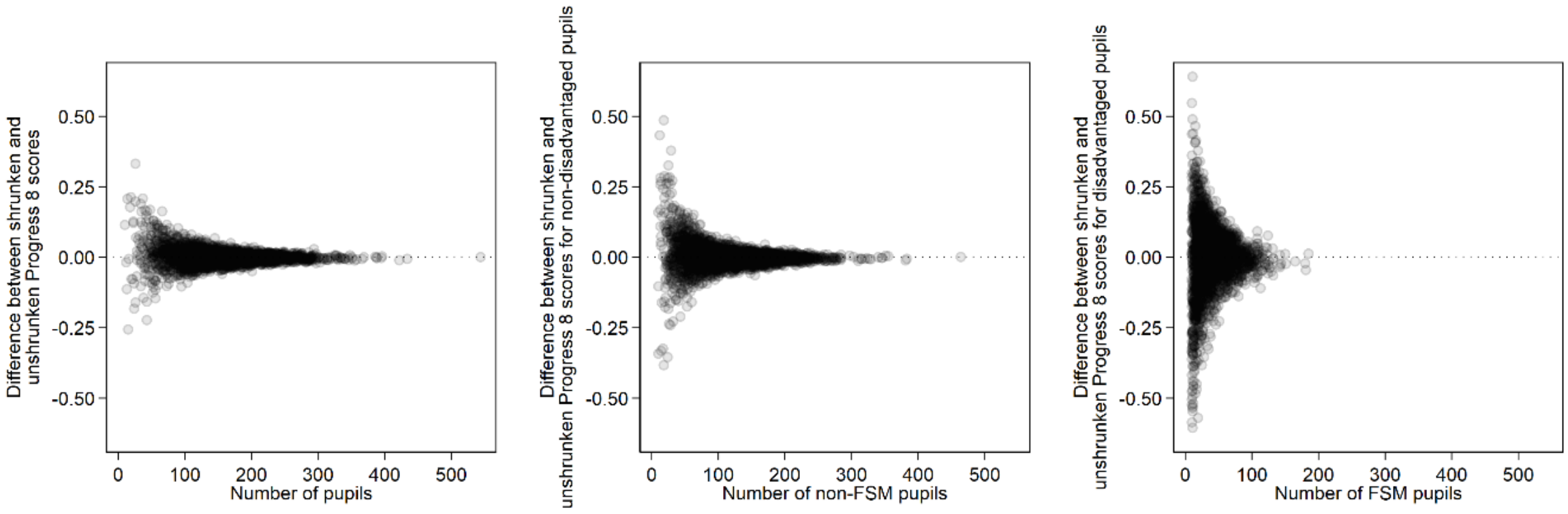

Figure 6.

Scatterplot of predicted pupil Attainment 8 scores (top) and predicted pupil Progress 8 scores (bottom) against KS2 score, plotted separately by linear regression model used, ‘single-level’ or multilevel.

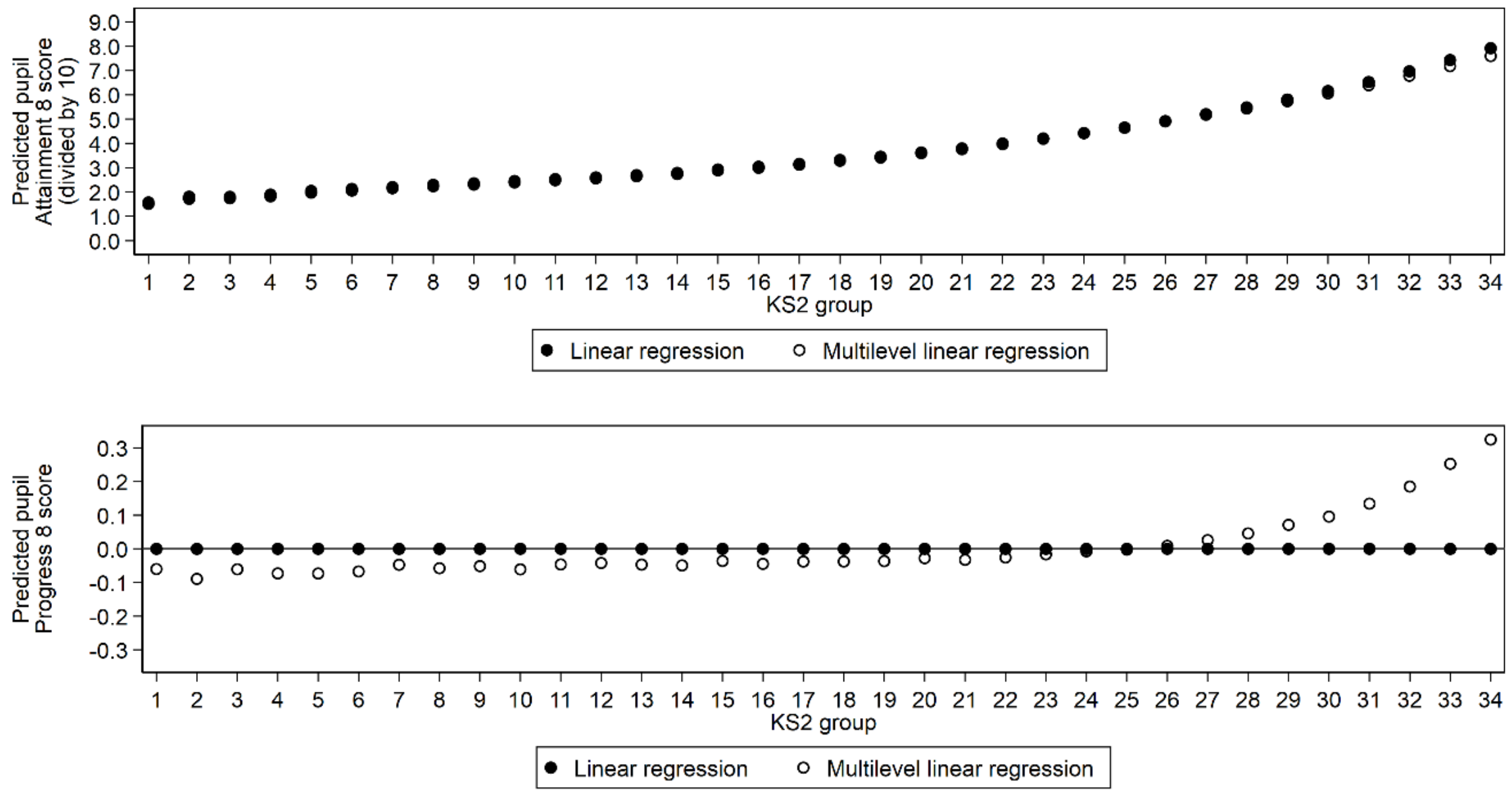

Figure 7.

Scatterplot of change in school Progress 8 score associated with a shift from using a ‘single-level’ linear regression model to a multilevel linear regression model plotted against school mean KS2 score.

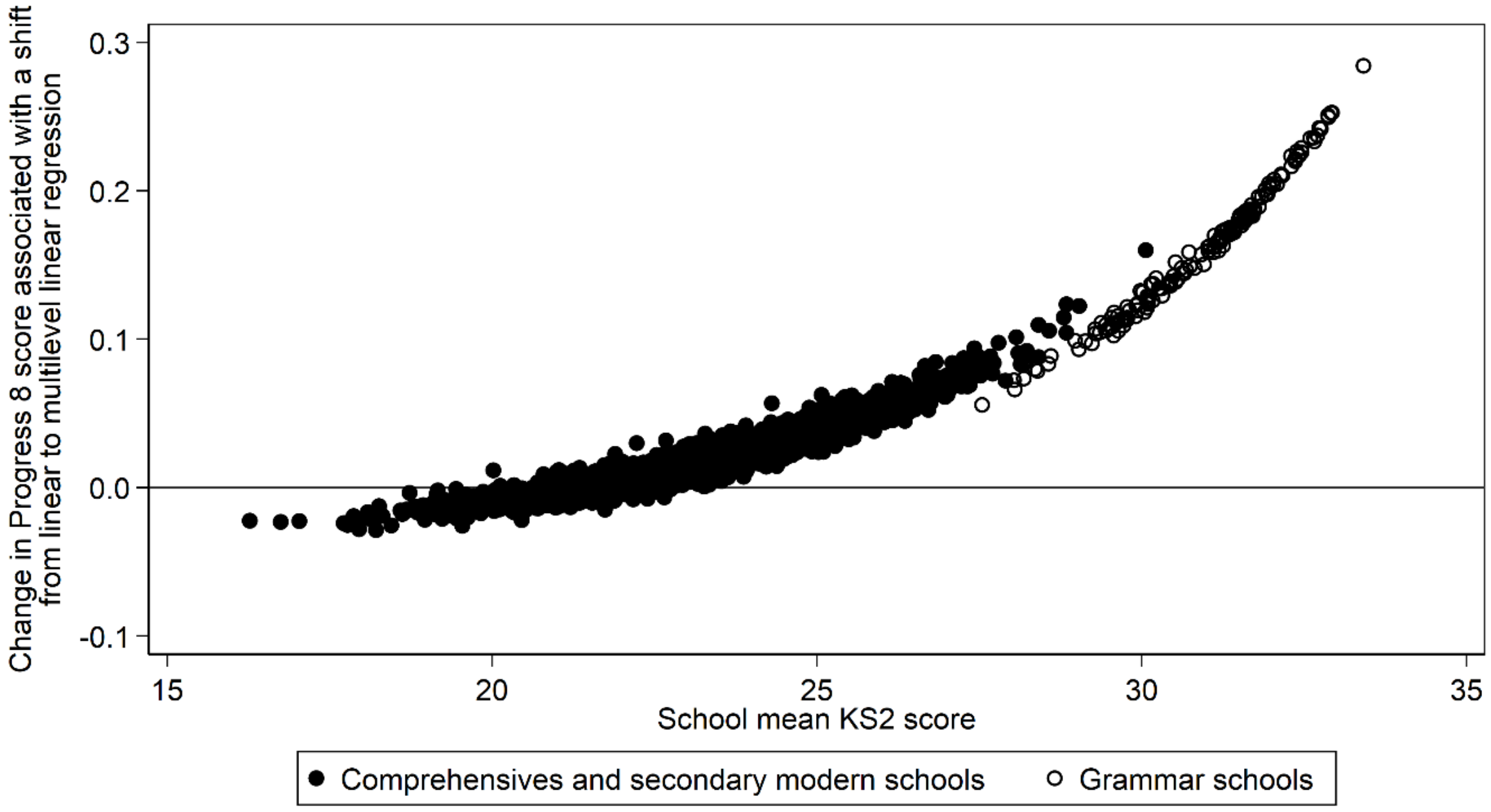

Figure 8.

Progress 8 scores and 95% confidence intervals over time and an overall three-year average for two schools in Bristol.

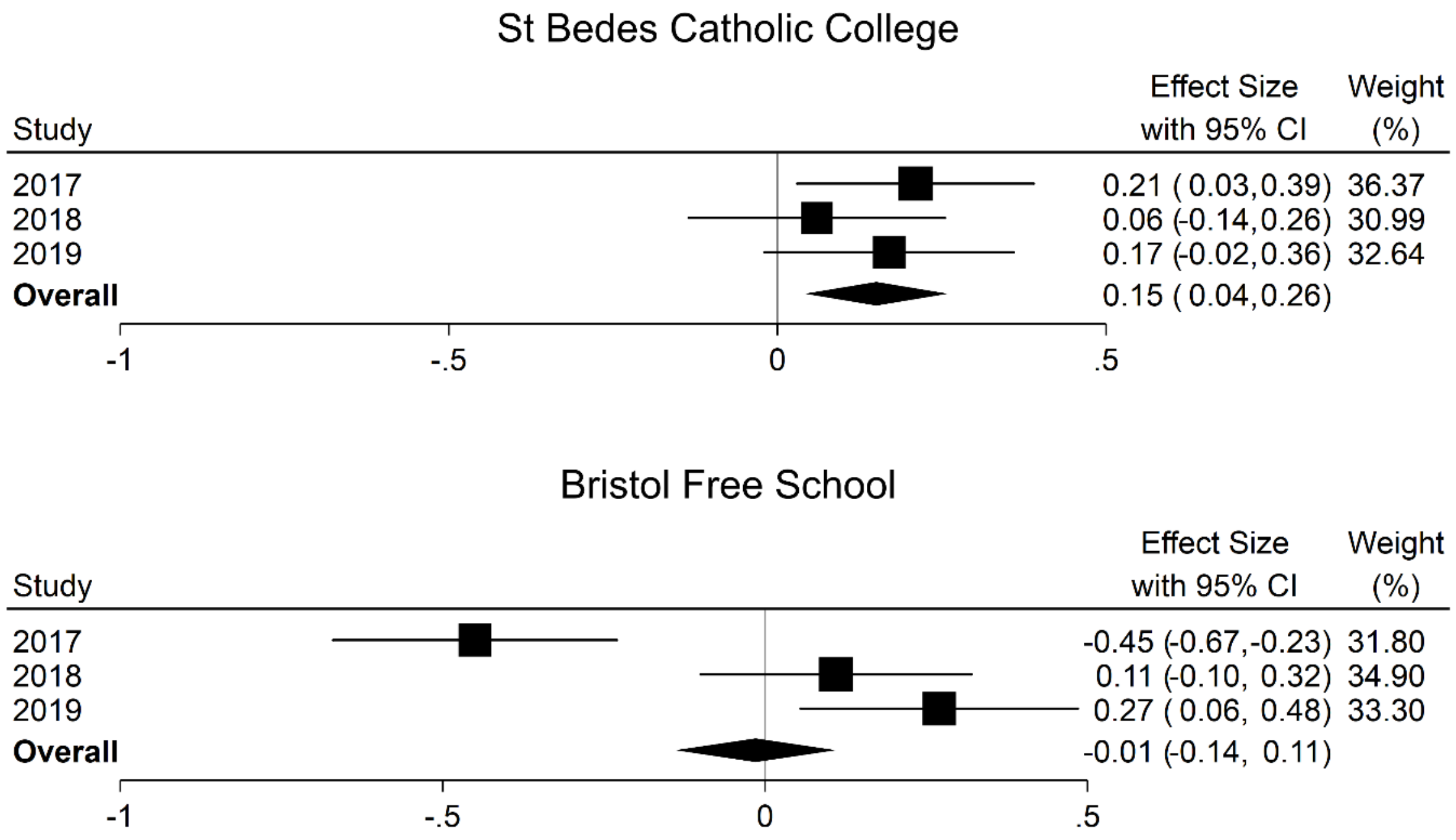

## Supplementary Material

In this section we provide technical information on the calculation of a school's Progress 8 score (Department for Education, 2020a), including how the Progress 8 methodology can be formulated as an application of linear regression.

Let $y_{ij}$ and $x_{ij}$ denote the Attainment 8 and KS2 score of pupil $i$ ($i = 1, \ldots, n_j$) in school $j$ ($j = 1, \ldots, J$). A pupil's Progress 8 score is calculated as the difference between their Attainment 8 score and the average Attainment 8 scores among all pupils nationally who had the same prior attainment as measured by pupils' average KS2 test scores in reading and maths. This difference is divided by 10 as school Progress 8 is reported on a per subject basis (recall that while Progress 8 covers eight qualifications, English and Maths GCSE grades are double counted). Let $p_{ij}$ denote the resulting pupil Progress 8 score.

The school Progress 8 score $\bar{p}_{.j}$ is then calculated as the average pupil Progress 8 score in each school.

$$\bar{p}_{.j} = \frac{1}{n_j} \sum_{i=1}^{n_j} p_{ij} \qquad (1)$$

From 2018 onwards, the DfE limited how negative a pupil's progress 8 score could be to help stop distortion of the overall performance of a school. Minimum thresholds are set for each prior attainment group, and pupils whose scores are below these thresholds have their scores altered to be set at the minimum. The threshold is determined by a set number of standard deviations below the mean for each prior attainment group so that approximately 1% of pupils are identified for adjustment nationally.

The lower and upper limits of the 95% confidence intervals for the school Progress 8 scores are calculated as

$$\widehat{\mathrm{LL}}(\hat{p}_j) = \bar{p}_{.j} - 1.96 \times \widehat{\mathrm{SE}}(\bar{p}_{.j}) \tag{2}$$

$$\widehat{\mathrm{UL}}(\hat{p}_j) = \bar{p}_{.j} + 1.96 \times \widehat{\mathrm{SE}}(\bar{p}_{.j}) \tag{3}$$

where 1.96 is the critical value for the 95% confidence intervals and $\widehat{\mathrm{SE}}(\bar{p}_{.j})$ is the standard error associated with the school Progress 8 score. The latter is calculated as

$$\widehat{\mathrm{SE}}(\hat{p}_j) = \frac{\hat{\sigma}}{\sqrt{n_j}} \tag{4}$$

where $\hat{\sigma}$ donates the standard deviation of pupils' Progress 8 scores nationally

$$\hat{\sigma} = \sqrt{\frac{\sum_{i=1}^{N}(p_{ij} - \bar{p}_{..})^2}{N-1}} \tag{5}$$

Progress 8 bandings are then calculated as follows

| Banding | Definition | |
|---|---|---|
| | Score | Significant |
| 5 = Well above average | $0.5 \leq \bar{p}_{.j}$ | Yes if $\widehat{\mathrm{LL}}(\hat{p}_j) > 0$ |
| 4 = Above average | $0 < \bar{p}_{.j} < 0.5$ | Yes if $\widehat{\mathrm{LL}}(\hat{p}_j) > 0$ |
| 3 = Average | Any | No if $\widehat{\mathrm{LL}}(\hat{p}_j) < 0$ &$\widehat{\mathrm{UL}}(\hat{p}_j) > 0$ |
| 2 = Below average | $0.5 < \bar{p}_{.j} < 0$ | Yes if $\widehat{\mathrm{UL}}(\hat{p}_j) < 0$ |
| 1 = Well below average | $\bar{p}_{.j} \leq -0.5$ | Yes if $\widehat{\mathrm{UL}}(\hat{p}_j) < 0$ |

We note in passing two oddities with the way the 95% confidence intervals are calculated. First, the use of 1.96 for the 95% confidence intervals implicitly assumes that school cohorts are very large whereas they are rather small (the average school cohort consists of approximately 160 pupils). Correcting for this would slightly widen the 95% confidence intervals, especially for small schools. For example, while 160 pupils implies a multiplier of 1.97, just 10 pupils implies a multiplier of 2.23. Second, the definition of $\hat{\sigma}$ is the standard deviation of pupil Progress 8 scores nationally whereas a more usual choice would be to define it as the standard deviation of pupil Progress 8 scores within the school under consideration, or perhaps the average across all within-school standard deviations. The latter is smaller than the standard deviation of pupil Progress 8 scores nationally. Correcting for this would slightly narrow the 95% confidence intervals. As the two corrections work in opposite directions, they will to some extent cancel each other out.

In Leckie and Goldstein (2019) we noted that we can formulate the Progress 8 methodology as an application of conventional linear regression. Namely a linear regression of $y_{ij}$ on $x_{ij}$ where we enter $x_{ij}$ as a series of 34 dummy variables $x_{1ij}, \dots, x_{34ij}$, one for each unique value of $x_{ij}$ and where we have omitted the usual constant term. This is an important insight as most school value-added performance measures are instead derived using multilevel linear regression and the two approaches lead to different estimates of school performance and in some scenarios these differences may be qualitatively important (see Section 6). A further benefit of formulating the Progress 8 methodology as an application of conventional linear regression is that we can then apply any of the usual extensions possible in linear regression, for example, it is easy to see how adjustments can be made for pupil demographic and socioeconomic background characteristics and their interactions simply by adding them as additional covariates (Section 4).

The linear regression can be written as

$$y_{ij} = \beta_1 x_{1ij} + \cdots + \beta_{34} x_{34ij} + r_{ij} \tag{6}$$

The regression coefficients $\beta_1, \ldots, \beta_{34}$ measure the expected value of $y_{ij}$ for each value of $x_{ij}$. In other words, the national average Attainment 8 score for pupils at each observed value of KS2 prior attainment.

Fitting this model by ordinary least-squares allows us to estimate these values $\hat{\beta}_1, \ldots, \hat{\beta}_{34}$. We can then assign these values to pupils via the usual linear regression prediction equation

$$\hat{y}_{ij} = \hat{\beta}_1 x_{1ij} + \cdots + \hat{\beta}_{34} x_{34ij} \quad (7)$$

where $\hat{y}_{ij}$ denotes the predicted Attainment 8 score for each pupil (the average Attainment 8 scores among all pupils nationally who had the same prior attainment).

The pupil Progress 8 score is then calculated as the difference between each pupil's actual Attainment 8 score $y_{ij}$ and their predicted Attainment 8 scores $\hat{y}_{ij}$. This difference is as before divided by 10. The pupil Progress 8 score $p_{ij}$ is therefore equal to the predicted residual from the above linear regression $\hat{r}_{ij}$ divided by 10. The school Progress 8 scores are once again the school averages of the pupil Progress 8 scores.

From 2018 onwards, the DfE limited how negative a pupil's progress 8 score could be to help stop distortion of the overall performance of a school. Minimum thresholds are set for each prior attainment group, and pupils whose scores are below these thresholds have their scores altered to be set at the minimum. The threshold is determined by a set number of standard deviations below the mean for each prior attainment group so that approximately 1% of pupils are identified for adjustment nationally.

Further information on Progress 8 is provided in the following documents: